\documentstyle[12pt]{article}
\newcommand{\ket}[1]{\mbox{$| #1 >$}}
\newcommand{\bpt}{\mbox{\boldmath $p$}_{T}}

\newcommand{\bP}{\mbox{\boldmath $P$}}

\newcommand{\bS}{\mbox{\boldmath $S$}}
\begin{document}

\bigskip

\bigskip
\begin{center}
{\Large \bf  The Physics of ELFE.\footnote {presented at the 7th Journ\'ees
 d'Etudes SATURNE, Ramatuelle (1996)}}

\bigskip
 {\bf Bernard Pire}
\end{center}
{\it
\hspace{1cm}
Centre de Physique Th\'eorique\footnote{  Centre de Physique Th\'eorique
 is unit\'e propre du CNRS.}, Ecole Polytechnique, F91128 Palaiseau, France}

\vspace{1cm}

\abstract{This paper presents an overview of the physics program of  the
15-30~GeV continuous beam electron facility proposed by the European
community of nuclear physicists  to study the quark and gluon structure of
hadrons.  The goal of  this new
facility  is  to  explore  the  quark  structure  of matter by exclusive
and semi-inclusive electron scattering from nuclear targets.}

\section{Introduction}

Recently,  the Nuclear Physics European  Collaboration Committee (NuPECC)
of the European  Science Foundation  has  recommended~\cite{NUPE95}  the
 construction  of  a  15-30 GeV  high intensity continuous beam  electron
 accelerator.   The goal of  this new facility  is  to  explore  the  quark
 structure  of matter by exclusive and semi-inclusive electron scattering from
 nuclear targets, since although one knows the  microscopic theory for the
 strong interactions, which is the quantum chromodynamics of colored quarks and
 gluons~\cite{QCD20}, {\em one does not understand how quarks build up hadrons}.

The ELFE  research  program\cite{ELFE} lies  at  the  border
  of  nuclear and particle physics. Most  of the predictions  of QCD are
 only valid at  very high energies  where  perturbation  theory  can  be
 applied.  Understanding however how  hadrons  are  built, is  the domain of
confinement where  the coupling  is strong.   Up  to now  there are only
crude theoretical  models of  hadronic structure  inspired by  QCD.  One
hopes that in the next  ten years major developments of  nonperturbative
theoretical methods such as lattice gauge theory will bring a wealth  of
results on the transition  from quark to hadron.   However, many theorists think
 that it is fundamental  to guide   theory   with   accurate, quantitative  and
 interpretable measurements obtained by electron scattering experiments.

The research program of ELFE addresses the questions raised by the quark
structure of matter: the role  of quark exchange,  color transparency,
flavor  and  spin  dependence  of  structure  functions  and differences
between  quark   distributions  in   the  nucleon   and  nuclei,   color
neutralization in the hadronization of a quark\ldots All these questions
are some of the many exciting facets of the fundamental question:

\begin{center}
{\bf ``How do color forces build up hadrons from quarks and gluons? ''}
\end{center}
\noindent
ELFE will focus on the following research topics:
\begin{itemize}
\item {\bf Hadron structure} as revealed by  hard exclusive reactions :
 baryon form factors,   real and virtual Compton scattering,
electro and photo-production of mesons ($\pi$,$K$,$\rho$,$\phi$ ...),
 meson ($\pi$,$K$, ...) form factors.
\item{\bf Evolution from  mini-hadron to hadron} in Color Transparency
experiments.
\item {\bf Vector mesons and heavy quarks}: diffractive production and exclusive
 scattering at high transfers.
\item {\bf Space time picture of quark hadronisation} through the study of
absorption by nuclear medium.
\item {\bf Separation of  Valence and Sea } content of the proton
  by tagged structure functions measurements.
\item {\bf Study of  spin structure } of the nucleon through semi-inclusive
experiments.
\item {\bf Light nuclei short distance structure } through form factor
measurements and deep inelastic scattering at $x>1$.
\end{itemize}

\section{\bf Hard Exclusive reactions:  A new tool}

Exclusive reactions are processes in which the final state
is completely
resolved.  They  are important since  at high momentum  transfers
they allow to study what may be called the simplest non perturbative objects.
This is due to a remarkable factorization property of scattering matrix elements
into a long distance confinement controlled hadronic distribution amplitude
convoluted with a short distance perturbatively calculable quark hard scattering
amplitude.

One starts with the Fock expansion\cite{EXCLTHEORY} with a fixed number of
quarks and gluons for a proton state of momentum $P$ :
$$
\ket{P} =
\Psi^P_{qqq} \ket{qqq} +
\Psi^P_{qqq,g} \ket{qqq,g} +
\Psi^P_{qqq,q\bar{q}} \ket{qqq,q\bar{q}} +
 ...
$$
where
$
\psi^P_l(x_i,\vec{k}^{\perp}_{i})$
describes how the $l$ quarks and gluons share the proton momentum.
The    wave  functions  $\Psi^P_l$  are  functions  of light-cone
momentum fraction $x_i$, transverse momentum $k^\perp_i$ and helicities.
They contain  the  information  on  quark  confinement  dynamics.
Here the quarks are ``current'' quarks and not ``constituent'' ones.

Of particular interest is
$
\Psi^P_{qqq}
$,
 the valence proton wave function. This is the simplest non-perturbative
object which recalls the color singlet nature of a confined quark
configuration.

Let us take the example of the proton form factor. Dimensional arguments easily
show that the ``three quarks'' hard scattering amplitude gives a contribution
 proportional to $\left[1/{Q^2}\right]^2$ due to the two gluon
 propagators, whereas the ``3~quarks~1~gluon'' amplitude, requiring three gluon
 propagators, contributes for $\left[1/{Q^2}\right]^3$. The argument may be
repeated for more participating constituents and for any reactions.
Thus, the valence component $\Psi^P_{qqq}$ turns out to be the dominant one in
hard exclusive reactions.

Factorization is then the statement
 that a hard matrix element can be written as
\begin{equation}
{\cal {M}} \simeq \phi^{*P}_{qqq}  \otimes T_H^{qqq,qqq}
\otimes\phi^P_{qqq},\label{LEADING}
\end{equation}
for a reaction with one proton in the initial and in the final state,
up to $1/Q^2$ corrections. Integrals  over momentum fractions  $x_i$ and $y_j$
 are implicit. Here
$$
\phi^P_{qqq}(x_i) ={\int} [dk^\perp] \Psi^P_{qqq}(x_i,\vec{k}^{\perp}_{i})
$$
is the proton   Valence Distribution Amplitude,
and $T_H^{qqq,qqq}$ is a hard scattering amplitude, calculable in
perturbative QCD.

\begin{figure}
\begin{center}
\vspace{5cm}
Fig 1: The factorization of a hard scattering amplitude \label{hard_bw}
\end{center}
\end{figure}

The applicability of this factorization
in a definite energy domain can be tested through some definite statements, such
 as the logarithmically corrected dimensional counting rules, the helicity
conservation law and the appearance of color transparency.
 The few data available~\cite{gpm_slac} (see figure 2) indicate that the ELFE
parameters indeed correspond to this well defined physics domain. Nevertheless,
 checking factorization will be a necessary prelude of the experimental program
 at ELFE.
\begin{figure}
\begin{center}
\vspace{4cm}
Fig.2: $Q^2$ evolution of the proton form factor. Above 10 GeV$^2$,
scaling is established.  \label{gpm_slac}
\end{center}
\end{figure}

The analysis of QCD  radiative corrections to any exclusive amplitude  has
shown~\cite{EXCLTHEORY}  that the factorized distribution  amplitudes  obey  a
renormalization group equation, leading  to a well understood  evolution
in terms  of perturbative  QCD.   At asymptotic  $Q^2$, the distribution
amplitudes simplify, e.g. for the proton valence distribution amplitude:
\begin{equation}
\phi^P_{qqq}(x_1,x_2,x_3,Q^2)\rightarrow K x_1x_2x_3\delta (1-x_1-x_2-x_3)
\end{equation}
\noindent
The $Q^2$ evolution  is however sufficiently  slow for the  distribution
amplitude  to  retain  much   information  at  measurable  energies   on
confinement physics.  The experimental strategy of ELFE physics is  thus
to sort out  the hadron distribution  amplitudes from various  exclusive
reactions to learn about the dynamics of confinement.

At finite $Q^2$  the proton valence wave-function can be written as a series
 derived from the leading logarithmic analysis,  in terms of
 Appell polynomials $P_i(x_i)$, as~\cite{EXCLTHEORY}
\begin{equation}
\label{APPEL}
\begin{array}{c}
\phi^P_{qqq}(x_i,Q^2)= x_1 x_2 x_3 \delta (1-x_1-x_2-x_3)
\left\{\right.
\left({\alpha_S(Q^2)\over\alpha_S(Q_0^2)}\right)^{\lambda_0}  A_0+
{{21}\over{2}}
\left({\alpha_S(Q^2)\over\alpha_S(Q_0^2)}\right)^{\lambda_1} A_1 P_1(x_i)\\
+{{7}\over{2}}
\left({\alpha_S(Q^2)\over\alpha_S(Q_0^2)}\right)^{\lambda_2} A_2 P_2(x_i)
+...\left.\right\}
\end{array}
\end{equation}
with $\lambda_0 = {{2} \over {27}} < \lambda_1 < \lambda_2 \dots$ The
 unknown coefficients $A_i$ are governed by confinement dynamics.

Due to the  smallness of exclusive amplitudes at  large transfers,
existing  high  energy   electron accelerators, designed to study electroweak
 physics, cannot  give access to  these distribution amplitudes.  The only  way
to  study  exclusive  reactions  at   large transfer is  to use  a dedicated
high intensity continuous beam  accelerator.

Photo- and electro-production of mesons at large angle will enable to
probe $\pi$ and $\rho$ distribution amplitudes in  the same way. The
production of $ K \Lambda $ final states will enable to explore strange
 quark production, for which the diagrams in the hard process are more
restricted. Not
much theoretical analysis of these possibilities has however been worked
out except under the simplifying assumptions of the diquark
model~\cite{DIQUARK}.

\section{Color Transparency}

 Configurations of small transverse extension
are selected by hard exclusive reactions. Indeed, in the Breit frame
 where the virtual photon is collinear to the incoming proton which flips
 its momentum, the first hit quark changes its direction and gets a momentum
$O(Q)$; it must transmit this information to its comovers within its light cone;
 this can only be achieved if the transverse separation is smaller than
 $O(1/Q)$. This is  the basis of the Color Transparency phenomenon~\cite{jpr}.

Hard exclusive scattering ( with a typical large $Q^2$ scale)  selects a very
  special quark configuration: the minimal valence state where
all quarks are  close together,  forming a  small size  color
neutral  configuration  sometimes  referred  to  as a {\em mini hadron}.
This mini hadron  is not a  stationary state and  evolves to build  up a
normal hadron.

Such  a color  singlet system  cannot emit  or absorb  soft gluons
which carry energy or momentum smaller than $Q$.  This is because  gluon
radiation --- like photon radiation in QED --- is a coherent process and
there is thus destructive interference between gluon emission amplitudes
by quarks  with ``opposite''  color.   Even without  knowing exactly how
exchanges  of   soft  gluons   and  other   constituents  create  strong
interactions, we  know that  these interactions  must be  turned off for
small color singlet objects. Letting the mini-state evolve during its travel
through different nuclei
of various  sizes allows  an indirect  but unique  way to  test how  the
squeezed mini-state  goes back  to its  full size  and complexity,  {\em
i.e.} how  quarks inside  the proton  rearrange themselves  spatially to
``reconstruct'' a normal size hadron.   In this respect the  observation
of baryonic resonance  production as well  as detailed spin  studies are
mandatory.

To the extent that hard scattering reactions in free space are understood as  a
function of $Q^2$, experiments on nuclei will allow to measure the color
screening properties of QCD. The quantity to be measured  is the transparency
 ratio $T_r$ defined as:
\begin{equation}
T_r = \frac{\sigma_{\rm Nucleus}}{Z \sigma_{\rm Nucleon}} ~~~~~~~ or ~~~~~~~
\frac{\sigma_{\rm Nucleus}}{A \sigma_{\rm Nucleon}}
\end{equation}

The gauge nature of QCD leads to the prediction that $T_r \rightarrow 1 $ as
$Q^2$ grows. At large values  of $Q^2$, dimensional estimates  suggest
that $T_r$ scales as a function of $A^{\frac{1}{3}}/Q^2$~\cite{PR91}.
The  approach to the scaling behavior as well as  the value of $T_r$ as a
function  of the  scaling  variable  determine  the  evolution  from  the
pointlike configuration to the  complete hadron. This
effect can be measured in an $(e , e' p)$ reaction that provides
the best chance for a {\it quantitative} interpretation, and in vector meson
diffractive electroproduction as recently observed at Fermilab (see below).

\section{Vector Meson production}

Diffractive vector meson production is usually understood in terms of Pomeron
 exchange. The nature of the Pomeron is however quite subtle.
Recent theoretical and experimental progresses have strengthened the
case for a detailed study in different energy ranges; in the very high
energy domain of HERA, it has been shown\cite{diff} that the vector meson
diffractive electroproduction amplitude at high $Q^2$ was calculable in
terms of the
gluon structure function and of the meson distribution amplitude. In the fixed
angle regime (when $\rho$ is produced at large transfer and the energy is not
 much higher), the hard exclusive reaction framework should apply. The case of
 heavy quarkonia ($\Phi$ and $J/\Psi$) is
particularly interesting but demands quite high luminosities. In the medium
energy
range, ELFE will allow a detailed analysis of diffractive as well as high $p_T$
electroproduction at various $Q^2$ and with controlled inelasticity.

Using nuclear targets will enable to scrutinize the formation and propagation
of the final state. The phenomenon of color transparency should also be at work
here. The data (Fig.3) on diffractive electroproduction  of $\rho$ at
 high energy~\cite{Fermilab}{} indeed show
an interesting increase of the transparency ratio for $Q^2 \simeq 7 GeV^2$.
Since the initial lepton energy is around
$E \simeq 200 GeV$ the boost is high, which yields a short travel time for the
produced mini-hadron inside the nucleus; one should however note that it is very
 difficult to disentangle diffractive from inelastic events in this experiment,
which is of crucial importance to correctly define the transparency ratio.
\begin{figure}
\vspace{3in}
\centerline{\small Fig.3: The Transparency ratio as measured at
 FNAL\cite{Fermilab}}
\centerline{\small in diffractive electroproduction of $\rho$}
\end{figure}
 Moreover gluon exchange may open the possibility to induce color
rearrangements between quark clusters in nuclei in order to study "hidden color"
 components such as a color octet-color octet component in the deuteron.

The study of polarized $\Lambda$ and open charm electroproduction will bridge
the gap between the physics of the heavy quark sector and the physics of
hadronization.

\section{Semi-inclusive reactions and hadronization}

Most of the time, the quark scattered by a large $Q^2$ virtual photon leads to
a complex multiparticle final state through the process of hadronization.
These events are better analized through an inclusive or semi-inclusive
formalism.
 The fully inclusive case has already been much studied. Up to order $1/Q$ and
including polarization, the cross section may be written as~\cite{Mulders}
\begin{eqnarray}
\frac{d\sigma} {dx\,dy} & = &
\frac{4\pi \alpha^2\,s}{Q^4}\ \Biggl\{\,
         \left\lgroup \frac{y^2}{2}+1-y\right\rgroup x f_1(x)
         + \lambda_e\,\lambda
          \,y \left\lgroup 1-\frac{y}{2} \right\rgroup x g_1(x)
\nonumber
\\&& - \lambda_e\,\vert \bS_\perp\vert\,\frac{M}{Q}
            \,2\,y\sqrt{1-y}\, \cos (\phi_s)\,x^2\, g_T(x)\,\Biggr\} .
\end{eqnarray}
with the usual scaling variables $x = Q^2/2P\cdot q$ and $y = P\cdot
k/P\cdot q$,
and the sum over quark flavors is understood.
The twist-3 function surviving after $k_T$-integration,
 $g_T(x) $, appears at subleading order.

ELFE will mostly contribute to inclusive studies  in two specific directions:

\noindent
- the $x > 1$ region where exotic phenomena - such as the existence of six-quark
structures in the nucleus - reveal themselves.

\noindent
- the higher twist components may be separated by careful use of polarization
asymmetries; this will help to understand quark-gluon correlations in the
nucleon
and nuclei.

\bigskip

Semi-inclusive scattering studies have not been much developped up to present
work with the Hermes detector at HERA. For these observables, one defines
the scaled hadron energy  $z = { E_{p'} \over \nu} $, where $\nu$ is the virtual
 photon energy and $ E_{p'}$ the produced hadron energy, and the invariant mass
squared $W^2$ of the hadronic final state $W^2 = m^2 +Q^2+2 m \nu$ (m is
the mass
of the initial hadron). Including a dependence on the transverse
momentum~\cite{Mulders}through a function $\cal G$ in the quark
distribution and
fragmentation functions, one gets
\begin{eqnarray}
&&\frac{d\sigma}{dx\,dy\,dz\,d^2\bP_{h\perp}} \ = \
\frac{4\pi \alpha^2\,s}{Q^4}\,\sum_{a,\bar a} e_a^2\,
\left\lgroup \frac{y^2}{2}+1-y\right\rgroup  \,x f_1^a(x)\,D^a_1(z)
\,\frac{{\cal G}(Q_T;R)}{z^2}
\nonumber \\ && \qquad \quad \mbox{}
-\frac{4\pi \alpha^2\,s}{Q^4}\,\lambda\,\sum_{a,\bar a} e_a^2\,
(1-y)\, \sin (2\phi_h)\,\frac{Q_T^2\,R^4}{M M_h\,R_H^2\,R_h^2}
\,x h_{1L}^{\perp\,a}(x) H_1^{\perp\,a}(z)
\,\frac{{\cal G}(Q_T;R)}{z^2}
\nonumber \\ && \qquad \quad \mbox{}
-\frac{4\pi \alpha^2\,s}{Q^4}\,\vert \bS_\perp\vert
\,\sum_{a,\bar a} e_a^2\,\Biggl\{
(1-y)\,\sin(\phi_h + \phi_s) \,\frac{Q_T\,R^2}{M_h\,R_h^2}
\,x h^a_1(x) H_1^{\perp\,a}(z)
\nonumber \\ && \qquad\qquad\qquad\quad
+(1-y) \,\sin(3\phi_h - \phi_s) \,
\frac{Q_T^3\,R^6}{2M^2M_h\,R_H^4\,R_h^2}
\,x h_{1T}^{\perp\,a}(x) H_1^{\perp\,a}(z)
\Biggr\}\,\frac{{\cal G}(Q_T;R)}{z^2}
\nonumber \\ && \qquad \quad \mbox{}
+\frac{4\pi \alpha^2\,s}{Q^4}\,\lambda_e\lambda\,\sum_{a,\bar a} e_a^2\,
 y\left(1-\frac{y}{2}\right) \,x\,g^a_{1L}(x)\,D^a_1(z)
\,\frac{{\cal G}(Q_T;R)}{z^2}
 \\ && \qquad \quad \mbox{}
+\frac{4\pi \alpha^2\,s}{Q^4}\,\vert \bS_\perp\vert
\,\sum_{a,\bar a} e_a^2\,
y\left(1-\frac{y}{2}\right)\,\cos (\phi_h - \phi_s)\,\frac{Q_T\,R^2}{M\,R_H^2}
\,g^a_{1T}(x) D^a_1(z)
\,\frac{{\cal G}(Q_T;R)}{z^2} \nonumber .
\end{eqnarray}
All six twist-two $x$- and $\bpt$-dependent quark distribution functions for
a spin 1/2 hadron can thus be accessed in leading order asymmetries with
polarized lepton and hadron. One of the asymmetries involves
the transverse spin distribution $h_1^a$ \cite{Collins}.
On the production side, only two
different fragmentation functions are involved, the familiar unpolarized
fragmentation function $D_1^a$ and the "transverse spin" fragmentation
function $H_1^{\perp a}$.

The semi-inclusive physics program of ELFE is twofold:

\noindent
- a better understanding of the proton content will be achieved with the
help of
tagged structure function measurements: the expected high luminosity and
particle
identification will allow a precise measurement of the strangeness or intrinsic
charm content at medium to high $x$. Polarization measurements will open
the study
of transverse spin~\cite{cpr} as well as helicity distributions, hopefully
separating valence and sea quark components. For instance, $\Delta_\perp q(x)$
may be measured in the semi-inclusive process
\begin{equation}
e^-\ N^{\uparrow} \to e^-\ \Lambda^{\uparrow} + X\,;
\end{equation}
since the polarization of the $\Lambda$  is proportional to
$\Delta_\perp q(x) \times \Delta_\perp D_{\Lambda^\uparrow/q}(z)$.
The second factor is the analysing power of the "quark polarimeter"
$q^\uparrow \to \Lambda^\uparrow + X$.

\noindent
- the use of nuclear targets will lead to the precise measurement of attenuation
ratio
$$
R_A (z, \nu) = ({{\frac {1} {\sigma_A} \frac {d\sigma_A} {dz} }) /
    ({\frac {1} {\sigma_D} \frac {d\sigma_D} {dz} }})
$$
of hadron yields on a nucleus A and on  deuteron D. Relating this piece of
information to the physics of the color neutralization process requires some
modelling. One first needs to take into account the loss of energy of the
scattered quark in the nuclear medium, presumably by gluon radiation in a
hopefully calculable way. After some characteristic time, a colorless wave
packet
is formed which further interacts strongly within the nucleus, leading
finally to
the detected hadron. The flavor, $\nu$ and $z$ dependence of $R_A$ will help
discriminating models based on different assumptions and scenarios.

\section{Accelerator and detectors}

The choice of the energy range of 15 to 30 GeV for the ELFE  accelerator
is fixed by a compromise between

\begin{itemize}

\item Hard electron-quark scattering:   one must have  sufficiently high
energy  and  momentum  transfer  to  describe  the  reaction in terms of
electron-quark scattering.  The high  energy corresponds to a very  fast
process where the struck quark  is quasi-free.  High momentum  transfers
are necessary to probe short distances.

\item  The smallness of the exclusive cross sections when the energy increases,
 as exemplified by the quark counting rules~\cite{EXCLTHEORY}.

\item The size of the nucleus as a femto-detector is of the right order of
magnitude provided the Lorentz boost is not so large that most of the physics
 happen outside. This prohibits very large energies, as exemplified by the
data obtained at$ O(200 GeV)$ at the Cern muon beam, which show negligible
nuclear
effects.

 \end{itemize}

Exclusive and semi-inclusive  experiments are at  the heart of  the ELFE
project.  To avoid a prohibitively large number of accidental coincident
events a high duty cycle  is imperative.  The ELFE  experimental program
also  requires  a  high   luminosity  because  of  the   relatively  low
probability of exclusive processes.  Finally a good energy resolution is
necessary to identify specific reaction channels.  A typical  experiment
at 15  GeV (quasielastic  scattering for  instance) needs  a beam energy
resolution of about  5 MeV. At  30 GeV the  proposed experiments require
only  to  separate  pion  emission.

Due  to  the  very  low  duty  cycle  available at SLAC and HERA (HERMES
program)  one  can  only  perform  with  these  accelerators   inclusive
experiments and a very limited set of exclusive experiments.

\begin{center}

{\em ELFE  will be  the first  high energy  electron beam  beyond 10 GeV
\\with both high intensity and high duty factor.}

\end{center}

 The design proposed in 1992 for the machine\cite{ELFE} consisted of a 1 km
 superconducting 5 GeV linac, with three recirculations .
A different design is now considered  which combines  a test  linac of
 30 GeV with 1 \% duty cycle for the future e$^+$e$^-$ collider (TESLA)
and the existing  HERA ring for stretching the pulse \cite{Bonin}.
The  various  components  of  the  ELFE experimental physics program put
different requirements on  the detection systems  that can be  satisfied
only  by  a  set  of  complementary  experimental  equipment.   The most
relevant detector features are  the acceptable luminosity, the  particle
multiplicity, the angular acceptance and the momentum resolution.   High
momentum   resolution   ($5   \times   10^{-4}$)   and  high  luminosity
($10^{38}$~nucleons/cm$^2$/s)  can  be  achieved  by  magnetic  focusing
spectrometers.   For semi-exclusive  or exclusive  experiments with more
than two  particles in  the final  state, the  largest possible  angular
acceptance  ($\sim  4  \pi$)  is  highly  desirable.    The  quality and
reliability of large acceptance detectors have improved substantially in
the last two decades.  The design of the ELFE large acceptance detectors
uses state of  the art developments  to achieve good  resolution and the
highest possible luminosity.

\section{CONCLUSIONS}

The ELFE  research program  lies at  the border  of nuclear and particle
physics.  Most  of the predictions  of QCD are  only valid at  very high
energies  where  perturbation  theory  can  be  applied.    In  order to
understand  how  hadrons  are  built,  however,  one has to go in the domain of
confinement where the  coupling is strong.   It is  fundamental to guide
theory  by  the  accurate,  quantitative  and interpretable measurements
obtained by electron scattering experiments, in particular in  exclusive
reactions.

 This research domain is  essentially a virgin territory.   There are
only scarce experimental data with poor statistics. This  lack
  of  data  explains  to  a  large  extent  the  slow  pace of
theoretical progress.   The situation can  considerably improve due  to
technical breakthroughs in electron accelerating techniques.  We believe
that future significant progress  in the understanding  of the
evolution from quarks to hadrons will be triggered by new information coming
from dedicated machines such as the ELFE project.

The goal of the ELFE research program, starting from the QCD  framework,
is to explore the coherent and quark confining QCD mechanisms underlying
the strong force.  It is not to test QCD in its perturbative regime, but
rather to use  the existing knowledge  of perturbative QCD  to determine
the reaction mechanism and access the hadron structure.

\vspace*{0.5cm} \noindent

{\em ELFE will use the tools that have been forged by twenty years of
research in QCD, to elucidate the central problem of color interaction:
color confinement and the quark and gluon structure of matter.}

\begin{center}
{\large\bf Aknowledgements}
\end{center}
I would like to acknowledge the many lively discussions on Elfe with many
 colleagues and especially Jacques Arvieux, Christian Cavata, Bernard Frois,
 Thierry Gousset, Pierre Guichon, Jean-Marc Laget, Thierry Pussieux and
John P.Ralston.

\end{document}